# Rakeness in the design of Analog-to-Information Conversion of Sparse and Localized Signals

Mauro Mangia, *Student Member, IEEE,* Riccardo Rovatti, *Fellow, IEEE,* Gianluca Setti, *Fellow, IEEE*

*Abstract*—Design of Random Modulation Pre-Integration systems based on the restricted-isometry property may be suboptimal when the energy of the signals to be acquired is not evenly distributed, i.e. when they are both sparse and *localized*.

To counter this, we introduce an additional design criterion, that we call *rakeness*, accounting for the amount of energy that the measurements capture from the signal to be acquired.

Hence, for localized signals a proper system tuning increases the rakeness as well as the average SNR of the samples used in its reconstruction. Yet, maximizing average SNR may go against the need of capturing all the components that are potentially non-zero in a sparse signal, i.e., against the restricted isometry requirement ensuring reconstructability.

What we propose is to administer the trade-off between rakeness and restricted isometry in a statistical way by laying down an optimization problem. The solution of such an optimization problem is the statistic of the process generating the random waveforms onto which the signal is projected to obtain the measurements.

The formal definition of such a problems is given as well as its solution for signals that are either localized in frequency or in more generic domain.

Sample applications, to ECG signals and small images of printed letters and numbers, show that rakeness-based design leads to non-negligible improvements in both cases.

## I. INTRODUCTION

This paper is about the application of some recently developed signal-processing techniques to the sensing of physical quantities, i.e., to their conversion into a sequence of samples that can be processed by an electronic system for the most diverse purposes.

Conventional approaches to this are based on the celebrated Shannon-Nyquist theorem stating that the sampling rate must be at least twice the highest frequency in the band of the signal (the so-called Nyquist frequency). This principle is the basis of almost all methods of acquisition used in nowadays audio and video consumer devices, in the processing of medical images, in the operation of radio receivers, etc;

*Compressed Sensing* (CS) is a recently introduced paradigm for the acquisition/sampling of signals that violates the Shannon-Nyquist theorem providing that additional (actually surprisingly broad) assumptions can be made.

A bird's eye view of CS shows that it is based on two general concepts: *sparsity*, which materializes the needed additional assumption, and *incoherence* between coordinate systems.

Sparsity expresses the idea that the information content of a signal can be much less than what is suggested by his bandwidth, or, for a discrete-time signal, that the number of its true degrees of freedom may be much smaller than its time length. Actually, many natural signals are sparse in the sense that they have a very compact representation when expressed with respect to a suitable reference system and are therefore susceptible to CS.

Incoherence extends the concept of duality between time and frequency. It is used to formalize the fact that when two domains are incoherent, objects that have a small representation in the first of them spread their energy over a wide support when seen from the point of view of the other domain.

It is evident that the first domain is best one when it comes to express and characterize the signal while the second is to be preferred for sensing operations since even few scattered measurements have a chance of capturing the signal energy. This is exactly what happens, for example, if we want to acquire a sinusoidal profile of unknown frequency. Since such a signal is extremely sparse in the frequency domain, the only two non-zero components of its spectral profile are incredibly effective in representing it. Yet, nobody would probe the frequency axis at few frequencies with the hope of coming across the one at which the signal is present and thus being able to recover the amplitude and phase. Instead, we know very well that only few samples in the time domain are enough to capture all the signal features.

Generalizing all this, CS architecture analyzes the target sparse signal by taking few measurements in the domain in which the energy is widespread and thus easy to collect. If this is done properly, the resulting samples can be subsequently processed by algorithmic means to reconstruct the small representation in the domain in which the signal is sparse.

The theoretical and practical machinery needed to do this in realistic conditions is being rapidly developed to arrive at acquisition mechanisms that can be labeled as Analog-to-Information (AI) converters [1]. In fact, once that the proper domain has been found in the form of a waveform basis along which the signals can be expressed as a linear combination with a small number of non-zero coefficients, the actual information being carried by the signals will be found in the positions and the magnitudes of those coefficients [2][3] that

G. Setti is with ENDIF - University of Ferrara - Via Saragat, 1 - 44100 Ferrara (ITALY)

R. Rovatti is with DEIS - University of Bologna, Viale Risorgimento, 2 - 40136 Bologna (ITALY)

All authors are also with ARCES - University of Bologna - Via Toffano, 2/2 - 40125 Bologna (ITALY)

email: mmangia@arces.unibo.it, gianluca.setti@unife.it, riccardo.rovatti@unibo.it

are the true target of the CS procedures.

## II. CS FOR LOCALIZED SIGNALS

The leveraging on sparsity has been recently paired [4][5] with another technique widely used by engineers to spot information content in signals, i.e. the uneven distribution of average energy along properly defined bases (that, in general, are different from those for which sparsity can be identified)[1]

In the following, we will indicate such an uneven distribution of energy with the term *localization* and we will observe that, in general, it provides a different a-priori information with respect to sparsity. As it is consequently naturally to expect, we will be able to show that these signal features allows improved sensing operations.

The key assumptions under which this may happen are that (i) measurements are taken by projecting the signal onto a proper set of waveforms whose cardinality is smaller than the dimensionality of the signal, and (ii) the overall effect of disturbances in the sensing process (thermal noise, quantization errors, etc.) can be modeled as a projection-independent error.

When (i) and (ii) hold, a noise-tolerant reconstruction of the sparse signal from a number of measurements that is smaller than its dimensionality is commonly achieved by designing the projection operator so that it is a *restricted isometry* (RI), i.e. it approximately preserves the length of the sparse signal to which it is applied so that the ratio between the norm of such signal and that of its projection falls within an interval $[\sqrt{1-\delta}, \sqrt{1+\delta}]$ where the RI constant $0 \leq \delta \leq 1$ should be as small as possible [2][3].

Roughly speaking this means that, if the measurements come from a RI, the original signal energy is not lost in the projection and, when acquisition error is added, the signal-to-noise ratio (SNR) of the samples remains high enough to perform reconstruction.

This approach and its pairing with localization can be intuitively explained with reference to a simplified, low-dimensionality setting in which the signal to acquire $a$ has three components $(a_0, a_1, a_2)$ and is sparse since only one of its components is non-vanishing in each realization. More formally we may assume that $a_j \neq 0$ with probability $p_j$ for $j = 0, 1, 2$ and, obviously, for $p_0 + p_1 + p_2 = 1$. Furthermore, when it is non-zero, the $j$-th component of the signal is a realization of a random variable with variance $\sigma_j^2$ for $j = 0, 1, 2$.

It is worth stressing that the possibility of $p_0\sigma_0^2 \neq p_1\sigma_1^2 \neq p_2\sigma_2^2$ implies that localization and sparsity are two separate concepts. In fact, though $a$ is sparse by construction, its average energy concentrates on the axis whose associated $p_j\sigma_j^2$ is larger and this concentration depends on the unbalance between the probability-variances products.

Since $a$ is sparse, it can be reconstructed by measuring its projection on a two-dimensional plane. To define it, refer to Figure 1-(a) and note that the generic projection plane passing through the origin defines an angle $\theta_j \in [0, \pi[$ with each axis

[1]A typical example is the class of band-pass signals, which are localized in the frequency domain, i.e., with respect to the Fourier basis.

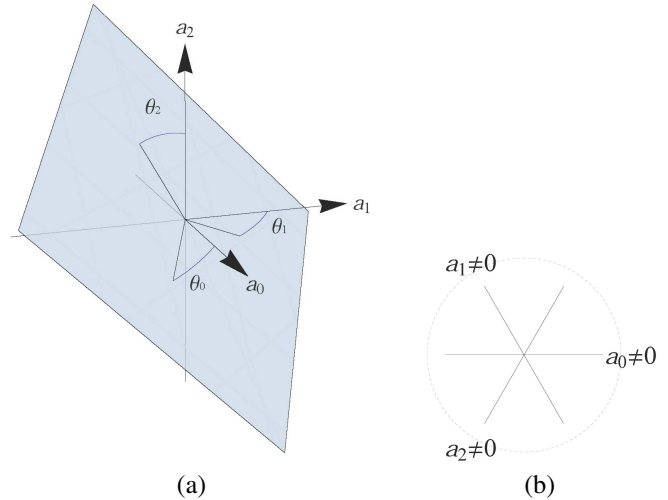

Fig. 1. A simple CS task using a projection plane designed by considering only the restricted isometry property (a). A graphical evaluation of the corresponding restricted isometry constant (b).

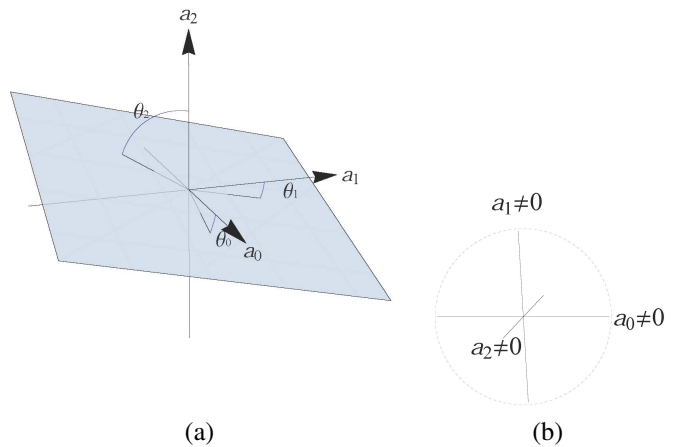

Fig. 2. A simple CS task using an optimized projection plane designed by merging rakeness and restricted isometry (a). A graphical evaluation of the corresponding restricted isometry constant (b).

$a_j$ for $j = 0, 1, 2$. These angles are such that $\sum_{j=0}^{2} \sin^2(\theta_j) = 1$.

Any set of angles $\theta_j \neq \pi/2$ for $j = 0, 1, 2$ is a feasible choice. This is shown in Figure 1-(b) that reports a unit radius circle on the projecting plane, along with the projections of unit-length segments centered in the origin and aligned with each of the three axis.

As long as the three projections have no point in common but the origin (so that, in general, the projections of the coordinate axes are distinct straight lines) the retrieval of the original signal in noiseless conditions can be ensured without complicated algorithms.

When noise comes into play, classical CS theory looks for planes corresponding to projection operators that are good RI. To do so, note that the ratio between the length of a segment aligned with the axis $a_j$ and that of its projection on the plane is $\cos(\theta_j)$. Hence, to minimize the RI constant we should choose each $\cos(\theta_j)$ as close as possible to 1, i.e. $\theta_0 = \theta_1 = \theta_2 = \sin^{-1}\sqrt{\frac{1}{3}}$ that is actually the case reported in Figure 1.

This choice clearly disregards the actual values of the probabilities $p_j$ and signal powers $\sigma_j^2$ for $j = 0, 1, 2$ and may be suboptimal.

To take these further information into account note that, since disturbances are introduced in acquiring the projections, they are independent from the plane. Hence, we may improve the SNR by choosing a plane that is able to *rake* a larger fraction of the signal power. We call this property *rakeness* and, in this case, to maximize it we have to maximize the power of the projection $\sigma^2 = p_0 \sigma_0^2 \cos^2(\theta_0) + p_1 \sigma_1^2 \cos^2(\theta_1) + p_2 \sigma_2^2 \cos^2(\theta_2)$.

With our assumption and by setting $\xi_j = \cos^2(\theta_j)$ for $j = 0, 1, 2$ this amounts to maximizing $\sigma^2 = p_0 \sigma_0^2 \xi_0 + p_1 \sigma_1^2 \xi_1 + p_2 \sigma_2^2 \xi_2$, subject to the constraint on the $\theta_j$ that becomes $\xi_0 + \xi_1 + \xi_2 = 2$. Assuming that $p_0 \sigma_0^2 > p_1 \sigma_1^2 > p_2 \sigma_2^2$, this criterion leads to $\xi_0 = \xi_1 = 1$ and $\xi_2 = 0$, i.e., a projections plane that coincides with the coordinate plane spanned by $a_0$ and $a_1$.

Clearly, the sheer maximization of the rakeness is not acceptable since any realization of $a$ in which $a_2 \neq 0$ would not be captured by the system or, in terms of the RI property, $\delta = 1$ since the $a_2$ axis belongs to the null-space of the projection operator.

This toy case highlights that RI and rakeness may be suboptimal as a design criterion when considered alone and that improvements may be sought addressing the trade-off between RI enforcement and rakeness maximization.

Such a trade off can be addressed both in a deterministic and in a statistical way.

Pursuing the deterministic path, one may choose a projecting plane like the one in Figure 2-(a) that still allows signals along $a_2$ to have a non-zero projection but clearly favors directions with the largest expected power. Figure 2-(b), that is analogous to Figure 1-(b) for the new plane choice, shows that this is detrimental in terms of the RI constant since the length of the projection of the segment along $a_2$ is substantially reduced. Yet, the lengths of the projections of the segments along $a_0$ and $a_1$ are increased and since these are the occurrences carrying more power on the average, the overall average acquisition quality may be improved.

The same improvement may be pursued in statistical terms by assuming that the projecting plane is chosen randomly at each measurement. In this case, the statistic of plane choices can be biased so that planes collecting larger energy are more probable, but planes allowing the acquisition of less important components are still possible.

This second setting is particularly interesting since random projections are already employed to guarantee good RI properties [6] and the main aim of this contribution is to show that the trade-off between RI and rakeness can be addressed by proper design of the statistical distribution of the projecting directions.

The rest of the paper is organized as follows. Section III will define the conversion architecture and lay down its mathematical model. Section IV introduces more formally the rakeness and its use as a design criterion. In doing this, to focus this exposition on application-oriented considerations, we accept that maximizing the energy of acquisitions is the right direction to go, thus postponing the statement of the formal chain of results starting from a mathematical definition of localization to a future contribution. This accepted, Section V describes a design path addressing the RI/rakeness trade-off when the signals to acquire are localized in the frequency domain, that is by far the most common domain for signal analysis. Section VI expands that view to include a generic adaptive domain, that is able to reveal localization in a large class of signals. Section VII shows how the theory developed in the two previous Sections can be applied to the acquisition of signals like ECGs and small images of printed letters and numbers. Some conclusions are drawn at the end, while a couple of lengthy derivations are reported in the Appendix.

### III. SYSTEM DEFINITION

We will concentrate on systems that perform AI conversion of sparse and localized signals by means of Random Modulation Pre-Integration (RMPI) [1].

This scheme sketched in Figure 3 acts on signals of the kind $a(t)$ where $t$ is most usually time but may also be any other indexing variable.

A "slice" of the signal $a$ (say for $-T/2 \leq t \leq T/2$ for some $T > 0$) is processed by multiplying it by a waveform $b(t)$ with a correspondingly sized support. The waveform $b(t)$ is made by amplitude modulated pulses whose modulating symbols are chosen from a certain set.

The most hardware friendly choices are rectangular pulses with binary ($\{0, 1\}$) or antipodal ($\{-1, +1\}$) symbols since multiplication can be implemented by a simple arrangement of switches that nicely embeds, for example, into switched-capacitor implementations [7].

The resulting waveform is then integrated or low-pass filtered to obtain a single value that is converted into a digital word by conventional means. Note that this further step also nicely fits into a switched-capacitor implementation that naturally manages charge integration.

Despite the fact that the rate (for time-indexed signals) or density (for generic signals) of the pulses may be very high and even larger than what a Nyquist-obeying acquisition would require, only the integrated values are actually converted.

This multiply-and-integrate operation materializes the scalar product $\langle a(t), b(t) \rangle$ and can be performed $M$ times (either serially or in parallel), each time considering a different waveform $b_j(t)$ ($j = 0, \ldots, M-1$) that is characterized by a set of modulating symbols drawn at random with a certain statistic.

The resulting projections $m_j = \langle a(t), b_j(t) \rangle$ for $j = 0, \ldots, M-1$ can be aligned in a measurement vector $\underline{m} = (m_0, \ldots, m_{M-1})^\top$. Figure 3 exemplifies the signals and the operations entailed by the acquisition of $m_j$ using antipodal PAM waveforms.

For what concerns signal reconstruction, we assume that $a$ is $K$-sparse, i.e., that there is a collection of $N$ waveforms

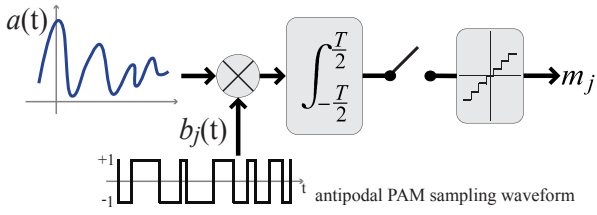

Fig. 3. Block diagram of RMPI architecture: the signal to acquire is multiplied by the $j$-th antipodal PAM waveform and fed into an integrator whose output is sampled and quantized to produce the digital conversion of the $j$-th measurement

$u_j(t)$ for $j = 0, \ldots, N-1$ such that every realization of $a$ can be written as

$$a(t) = \sum_{j=0}^{N-1} a_j u_j(t) \quad (1)$$

for certain coefficients $a_j$ such that at most $K < N$ of them can be non-zero at any time.

Plugging (1) into the definition of $m_j$ we get $m_j = \sum_{k=0}^{N-1} a_k \langle u_k(t), b_j(t) \rangle$. By defining the vector $\underline{a} = (a_0, \ldots, a_{N-1})^\top$, the $M \times N$ projection matrix $\underline{P} = [\underline{P}_{j,k}] = [\langle u_k(t), b_j(t) \rangle]$, and the vector $\underline{\nu} = (\nu_0, \ldots, \nu_{M-1})^\top$ accounting for the total noise affecting the projections, we have that

$$\underline{m} = \underline{P}\,\underline{a} + \underline{\nu} \quad (2)$$

is the reconstruction equality to be solved for the unknown $\underline{a}$ with the aid of its $K$-sparsity. In principle, this could be done by selecting, among all the vectors $\underline{a}$ satisfying (2), the one with the minimum number of non-zero entries. Since this is, in general, a problem subject to combinatorial explosion, many alternative theoretical and algorithmic methods have been developed allowing efficient and effective reconstructions [8][9][10]. Among all these possibilities, we will exploit the algorithm described in [9] in our experiments in Section VII.

Note that $\underline{\nu}$ takes into account at least the intrinsic thermal noise affecting the analog processing of $a(t)$ and the quantization noise due to digitalization. Since thermal noise is additive white and Gaussian (AWGN), its contribution to $\underline{\nu}$ is independent of the projecting waveforms $b_j(t)$ as long as they have constant energy. We assume that quantization noise is also approximately white and independent on the quantized input so that condition (ii) discussed in Section II is satisfied.

## IV. Restricted Isometry and Rakeness

To cope with the noise in $\underline{\nu}$, RI-based design [11] tries to make the RI constant $\delta$ of the projection operator as low as possible. This can be checked directly from the matrix $\underline{P}$. In fact, since the projection is applied to $K$-sparse vectors, we should consider each of the $\binom{N}{K}$ matrices $\underline{P}'$ that are built selecting $K$ of the $N$ columns of $\underline{P}$. If $\lambda_{\underline{P}'}^{\min}$ and $\lambda_{\underline{P}'}^{\max}$ are respectively the minimum and maximum among the singular values [12] of $\underline{P}'$ we have

$$\delta = \max_{\underline{P}'} \left\{ \max \left[ 1 - \lambda_{\underline{P}'}^{\min}, \lambda_{\underline{P}'}^{\max} - 1 \right] \right\}$$

To go further, we define the average rakeness $\rho$ between any two processes $\alpha$ and $\beta$ as

$$\rho(\alpha, \beta) = \kappa_\rho \mathbf{E}_{\alpha,\beta} \left[ |\langle \alpha, \beta \rangle|^2 \right] \quad (3)$$

where the constant $\kappa_\rho$ is used to switch the meaning of $\rho$ from "average *energy* of projections" ($\kappa_\rho = 1$) to "average *power* of projections" (e.g., $\kappa_\rho = T^{-1}$ for signals observed in $[-T/2, T/2]$).

It is worthwhile to highlight that $\rho(\alpha, \beta)$ depends on how the second-order features of the two processes combine. In fact, we may expand the definition as

$$\rho(\alpha, \beta) =$$
$$= \kappa_\rho \mathbf{E}_{\alpha,\beta} \left[ \int_{-T/2}^{T/2} \int_{-T/2}^{T/2} \alpha^*(t)\beta(t)\alpha(s)\beta^*(s) \mathrm{d}t\mathrm{d}s \right]$$
$$= \kappa_\rho \int_{-T/2}^{T/2} \int_{-T/2}^{T/2} \mathbf{E}_\alpha \left[ \alpha^*(t)\alpha(s) \right] \mathbf{E}_\beta \left[ \beta(t)\beta^*(s) \right] \mathrm{d}t\mathrm{d}s$$
$$= \kappa_\rho \int_{-T/2}^{T/2} \int_{-T/2}^{T/2} C_\alpha(t,s) C_\beta^*(t,s) \mathrm{d}t\mathrm{d}s \quad (4)$$

where $\cdot^*$ stands for complex conjugation and the two correlation functions $C_\alpha$ and $C_\beta$ have been implicitly defined.

From the toy example in Section II, we know that choosing the process $b$ that maximizes the rakeness $\rho(a,b)$ leads to good average SNR of the projections, but may destroy the RI property making the system insensitive to some signal components.

To counter this *over-tuning* effect one may require that the process $b$ is "random enough" to assign a non-zero probability to realizations that, despite being sub-optimal from the point of view of energy collection, allow the detection of components of the original signal that would be overlooked otherwise. Actually, this intuitive approach is fully supported by the existing results on the RI property. In fact, it is known [6] that if the matrix $\underline{P}$ is made of random independent entries, its RI constant is small with a substantially large probability.

In general, enforcing the randomness of a process can be a subtle task since the very definition of what is random (entropic, algorithmically complex, etc.) can be extremely sophisticated and also dependent more on philosophical than technical consideration.

Here, for simplicity's sake, we limit ourselves to energy/power considerations and define a measure of the (non)randomness of a process as its self-rakeness, i.e., the average amount of energy/power of the projection of one of its realization onto another realization when the two are drawn independently. The rationale behind this quantification of randomness is that, if $\rho(b,b)$ is high, then independent realizations of the process tend to align and thus to be substantially the same, implying a low "randomness" of the process itself [4].

This definition nicely fits into a mathematical formulation of the design path that increases the rakeness $\rho(a,b)$ while leaving $b$ random enough. In fact, given a certain sparse stochastic process $a$, we determine the stochastic process $b$ generating the projecting waveforms to employ in an RMPI architecture by solving the following optimization problem

$$\begin{aligned} \max_b \quad & \rho(a,b) \\ \text{s.t.} \quad & \langle b,b \rangle = e \\ & \rho(b,b) \leq re \end{aligned} \quad (5)$$

where $e$ is the energy of the projection waveforms and $r$ is a randomness-enforcing design parameter.

Roughly speaking, solving (5) will ensure that the resulting waveforms will have constant energy (due to constraint $<b,b>=e$) paired with the ability of maximizing the average SNR of the projections (thanks to the capability of maximizing the energy of the acquired samples since we impose that $\max_b \rho(a,b)$) while maintaining the chance of detecting components of the original signal that carry smaller amounts of energy/power (thanks to the fact that each realization of the process $b$ has $\rho(b,b) \leq re$, i.e., low autocorrelation and thus "large" randomness).

In (5), the parameter $e$ acts as a normalization factor, since if $b'$ is the solution for $e = e'$, then $b'' = \sqrt{e''/e'}b'$ is the solution of the same problem for $e = e''$.

On the contrary, $r$ is the parameter controlling the trade-off between the two design criteria we want to blend, i.e., RI and rakeness. Hence, different values of $r$ lead to waveform with different final performance.

Regrettably, though it is easy to accept that, thanks to their ability to maximize the energy of the samples, the resulting $b$ may be able to increase the performance of the overall sensing system, the latter may rely (especially in the reconstruction part) on heavily non-linear and iterative operations that are difficult to model. For this reason, though feasible bounds for the parameter $r$ can (and will) be derived theoretically in Section V and VI, the choice of its exact value is a matter of fine tuning of the global system, and it must be determined through numerical simulation.

## V. Localization in the Frequency Domain

In this Section we specialize (5) to the case in which the statistical features of $a$ that cause the localization of its energy/power can be straightforwardly highlighted by Fourier analysis.

We will concentrate on the time interval $[-T/2, T/2]$ and set $\kappa_\rho = T^{-1}$ in (3).

To express $\rho$ in terms of the frequency-domain features of the processes $\alpha$ and $\beta$ in it, let us assume that both of them are second-order stationary.

Leveraging on this, we may define the single-argument correlation functions $C_\alpha(s-t) = C_\alpha(t,s)$ and $C_\beta(s-t) = C_\beta(t,s)$ whose Fourier transforms are nothing but the power spectra $\hat{\alpha}(f)$ and $\hat{\beta}(f)$ of the two processes.

For $\rho(\alpha, \beta)$ we obtain

$$\begin{aligned} \rho(\alpha,\beta) &= \\ &= \frac{1}{2T} \int_{-T}^{T} C_\alpha(p) C_\beta^*(p) \int_{-T+|p|}^{T-|p|} \mathrm{d}q \mathrm{d}p \\ &= \int_{-T}^{T} C_\alpha(p) C_\beta^*(p) \left(1 - \frac{|p|}{T}\right) \mathrm{d}p \qquad (6) \\ &= \int_{-T}^{T} \int_{-\infty}^{\infty} \int_{-\infty}^{\infty} \hat{\alpha}(f) \hat{\beta}^*(g) e^{2\pi\mathrm{i}(f-g)p} \left(1 - \frac{|p|}{T}\right) \mathrm{d}p \mathrm{d}f \mathrm{d}g \\ &= \int_{-\infty}^{\infty} \int_{-\infty}^{\infty} \hat{\alpha}(f) \hat{\beta}^*(g) \int_{-T}^{T} e^{2\pi\mathrm{i}(f-g)p} \left(1 - \frac{|p|}{T}\right) \mathrm{d}p \mathrm{d}f \mathrm{d}g \\ &= \int_{-\infty}^{\infty} \int_{-\infty}^{\infty} \hat{\alpha}(f) \hat{\beta}^*(g) h_T(f-g) \mathrm{d}f \mathrm{d}g \qquad (7) \end{aligned}$$

where

$$h_T(f) = \int_{-T}^{T} e^{2\pi\mathrm{i}fp} \left(1 - \frac{|p|}{T}\right) \mathrm{d}p = \frac{\sin^2(\pi T f)}{\pi^2 T f^2}$$

For simplicity's sake we may focus on the antipodal case in which the projection waveforms have a constant-modulus amplitude ($\pm 1$), duration $T$, and thus automatically satisfy the constant energy constraint $<b,b>=e$ in (5) with $e = T$, needed to make projection tuning possible.

With this, the power spectrum of the projection waveforms can be designed by solving (5) re-expressed in the frequency domain. To do so, use (7) to rewrite $\rho(a,b)$ and $\rho(b,b)$ in (5) and consider

$$\begin{aligned} \max_{\hat{b}} \quad & \int_{-\infty}^{\infty} \int_{-\infty}^{\infty} \hat{a}(f) \hat{b}(g) h_T(f-g) \mathrm{d}f \mathrm{d}g \\ \text{s.t.} \quad & \int_{-\infty}^{\infty} \int_{-\infty}^{\infty} \hat{b}(f) \hat{b}(g) h_T(f-g) \mathrm{d}f \mathrm{d}g \leq rT \\ & \hat{b}(f) \geq 0 \qquad\qquad\qquad\qquad\qquad (8) \\ & \int_{-\infty}^{\infty} \hat{b}(f) \mathrm{d}f = 1 \\ & \hat{b}(f) = \hat{b}(-f) \end{aligned}$$

where the last three constraints encode the fact that $\hat{b}$ must be a power spectrum of a unit-power, real signal.

Once that $r$ is fixed, (8) can be solved by assuming that $\hat{a}$ concentrates its power in the frequency interval $[-B,B]$ and applying some kind of finite-elements methods, i.e., approximating all the entailed functions with linear combinations of basic function elements on which the integrals can be computed at least numerically.

As an example, select a frequency interval $[-B,B]$ and partition it $2n+1$ subintervals of equal length $\Delta f = 2B/(2n+1)$ $F_j = [j - \Delta f/2, j + \Delta f/2]$ for $j = -n, \ldots, n$. Assume now that $\hat{b}(f)$ is constant in each $F_j$ and define $\chi_j(f)$ as the indicator fucntion of $F_j$, i.e. $\chi_j(f) = 1$ if $f \in F_j$ and 0 otherwise. We have $\hat{b}(f) = \sum_{j=-n}^{n} b_j \chi_j(f)$ for certain coefficients $b_{-n}, \ldots, b_n$.

This can be substituted into (8) to obtain

$$\max_{b_{-n},\ldots,b_n} \sum_{j=-n}^{n} w_j b_j$$

$$\text{s.t.} \quad \begin{aligned} & \sum_{j=-n}^{n}\sum_{k=-n}^{n} b_j b_k W_{j,k} \leq rT \\ & b_j \geq 0 \quad j=-n,\ldots,n \\ & \Delta f \sum_{j=-n}^{n} b_j = 1 \\ & b_j = b_{-j} \quad j=-n,\ldots,n \end{aligned} \qquad (9)$$

with

$$w_j = \int_{-\infty}^{\infty}\int_{F_j} \hat{a}(f) h_T(f-g) \mathrm{d}f\mathrm{d}g$$

$$W_{j,k} = \int_{F_j}\int_{F_k} h_T(f-g) \mathrm{d}f\mathrm{d}g$$

This leaves us with the vector of $2n+1$ unknown coefficients $b_{-n},\ldots,b_n$ that must determined by solving an optimization problem characterized by a linear objective function and few linear and quadratic constraints. Plenty of numerical methods exist for solving such problems even for large number of basic-elements and thus for extremely effective approximations (commercial products such as MATLAB or CPLEX provide full support for large-scale version of these problems).

Once that the optimum $\hat{b}(f)$ has been computed, one may resort to known methods to generate an antipodal process with such a spectrum exploiting a linear probability feedback (LPF) [13][14][15]. Slices of length $T$ of this process can be used as projection waveforms in an RMPI architecture for the CS of the original $a$.

Note that, even if this is needed to arrive at a final working system, the core of rakness-based design concerns the solution of (5) for frequency-localized signals to obtain the best spectral profile of the projecting waveforms, independently of their physical realization. How such a spectral profile can be obtained using antipodal PAMs is an implementation-dependent choice, which allows to realize an hardware system for sparse and localized signal acquisition which does not require analog multipliers [5].

As far as the range in which $r$ should vary to administer the trade off between RI and rakeness, note that, since $\rho(b,b)$ is a measure of (non)randomness, it must be minimum when the process $b$ is white in its bandwidth, i.e., when $\hat{b}(f) = 1/(2B)$ for $f \in [-B, B]$ and $0$ otherwise.

Plugging this into (7) and defining $c = BT$ one gets

$$r \geq r^{\min} = \\ = \frac{\mathrm{Ci}(4\pi c) + 4\pi c \mathrm{Si}(4\pi c) - \log(4\pi c) + \cos(4\pi c) - \gamma - 1}{4\pi^2 c^2}$$

where $\gamma$ is the Euler's constant and Ci and Si are respectively the cos-integral and sin-integral functions.

The quantity $r^{\min} c$ is a monotonically and rapidly increasing function of $c$ with $\lim_{c \to \infty} r^{\min} c = 1/2$. Hence, we may safely use such an asymptotic value to set $1/(2c)$ as a suitable lower bound for $r$ in any practical conditions.

Again, from the meaning of $\rho(b,b)$ we got that it is maximum when the waveforms produced by the process are constant. This implies $C_b(\tau) = 1$ that can be plugged into (6) to obtain

$$r \leq r^{\max} = \frac{1}{T}\int_{-T}^{T}\left(1 - \frac{|p|}{T}\right) \mathrm{d}p = 1$$

Overall, the tuning of the overall system will optimize performance by choosing $r \in \left[\frac{1}{2c}, 1\right]$.

## VI. LOCALIZATION IN A GENERIC DOMAIN

Slices of second-order stationary processes (that enjoy a simple and well-studied characterization in the frequency domain) do not exhaust the set of signals that we may want to acquire.

To cope with more general cases assume to work in normalized conditions such that both the waveforms to be acquired and the projection waveforms have unit energy, i.e., $\int_{-\frac{T}{2}}^{\frac{T}{2}} |a(t)|^2 \mathrm{d}t = \int_{-\frac{T}{2}}^{\frac{T}{2}} |b(t)|^2 \mathrm{d}t = 1$, where the latter constrain sets $e = 1$ in (5).

When we comply with this assumption (possibly by scaling the original signals), if $C_x$ represents either $C_a$ or $C_b$, we have that

- $C_x$ is Hermitian, i.e., $C_x(t,s) = C_x^*(s,t)$;
- $C_x$ is positive semidefinite, i.e., for any integrable function $\xi(t)$ the quadratic form $\int_{-\frac{T}{2}}^{\frac{T}{2}}\int_{-\frac{T}{2}}^{\frac{T}{2}} \xi^*(t) C_x(t,s) \xi(s) \mathrm{d}t \mathrm{d}s = \mathbf{E}\left[\left|\int_{-\frac{T}{2}}^{\frac{T}{2}} x(t)\xi(t)\mathrm{d}t\right|^2\right]$ yields a non-negative result;
- $C_x$ has a unit trace, i.e.,

$$\int_{-\frac{T}{2}}^{\frac{T}{2}} C_x(t,t) \mathrm{d}t = \\ = \int_{-\frac{T}{2}}^{\frac{T}{2}} \mathbf{E}[|x(t)|^2] \mathrm{d}t = \mathbf{E}\left[\int_{-\frac{T}{2}}^{\frac{T}{2}} |x(t)|^2 \mathrm{d}t\right] = 1.$$

From this, we know (see e.g. [16]) that two sequences of orthonormal functions $\theta_0(t), \theta_1(t), \ldots$ and $\phi_0(t), \phi_1(t), \ldots$ exist, along with the sequences of real non-negative numbers $\mu_0 \geq \mu_1 \geq \ldots$ and $\lambda_0 \geq \lambda_1 \geq \ldots$ such that $\sum_{j=0}^{\infty} \mu_j = \sum_{j=0}^{\infty} \lambda_j = 1$ and

$$C_a(t,s) = \sum_{j=0}^{\infty} \mu_j \theta_j^*(t) \theta_j(s) \qquad (10)$$

$$C_b(t,s) = \sum_{j=0}^{\infty} \lambda_j \phi_j^*(t) \phi_j(s) \qquad (11)$$

By substituting the generalized spectral expansions for the two correlation functions (10) and (11) into (4) one gets

$$\rho(a,b) = \sum_{j=0}^{\infty}\sum_{k=0}^{\infty} \lambda_j \mu_k \Xi_{j,k}$$

$$\rho(b,b) = \sum_{j=0}^{\infty} \lambda_j^2$$

where the real and nonnegative numbers

$$\Xi_{j,k} = \left| \int_{-\frac{T}{2}}^{\frac{T}{2}} \phi_j(t)\theta_k^*(t)\mathrm{d}t \right|^2$$

are the squared modulus of the projections of each $\phi_j$ on every $\theta_k$ (and viceversa).

The orthonormality of the $\theta_k$ guarantees that the sum of the squared modulus of the projections of $\phi_j$ must equal the squared length of $\phi_j$ itself and thus, since $\phi_j$ is normal, that $\sum_{j=0}^{\infty}\Xi_{j,k} = 1$. Conversely, from the fact that the $\phi_j$ are orthonormal we have also $\sum_{k=0}^{\infty}\Xi_{j,k} = 1$.

Hence, the optimization problem (5) can be rewritten in totally generic terms as

$$\max_{\lambda}\max_{\Xi} \quad \sum_{j=0}^{\infty}\sum_{k=0}^{\infty} \lambda_j \mu_k \Xi_{j,k}$$

$$\text{s.t.} \quad \begin{aligned} \lambda_j &\geq 0 \quad \forall j \\ \sum_{j=0}^{\infty} \lambda_j &= 1 \\ \sum_{j=0}^{\infty} \lambda_j^2 &\leq r \\ \Xi_{j,k} &\geq 0 \quad \forall j,k \\ \sum_{j=0}^{\infty}\Xi_{j,k} &= 1 \quad \forall k \\ \sum_{k=0}^{\infty}\Xi_{j,k} &= 1 \quad \forall j \end{aligned} \quad (12)$$

Note that the two max operators address separately the problem of finding an optimal basis ($\max_\Xi$) and then the optimal energy distribution over that basis ($\max_\lambda$).

As far as the range of $r$ is concerned, assume to know that $J$ is an integer such that $\lambda_j = 0$ for $j \geq J$. It can be easily seen that $\max \sum_{j=0}^{J-1} \lambda_j^2$ subject to the constraints $\lambda_j \geq 0$ and $\sum_{j=0}^{J-1} \lambda_j = 1$ is 1 and is attained when $\lambda_0 = 1$ and $\lambda_j = 0$ for $j > 0$. It is also easy to see that $\min \sum_{j=0}^{J-1} \lambda_j^2$ subject to the constraints $\lambda_j \geq 0$ and $\sum_{j=0}^{J-1} \lambda_j = 1$ is $1/J$ and is attained when $\lambda_j = 1/J$ for $j = 0, \ldots, J-1$. Hence, $r \in [1/J, 1]$.

In particular, the lower bound $r \geq 1/J$ rewritten as $rJ \geq 1$ can be read as a general rule of thumb, i.e., the more random the process that generates the projection waveforms, the larger the number of non-zero eigenvalues in the spectral expansion of its correlation function.

The solution of (12) is derived in the Appendix and depends on the two partial sums

$$\Sigma_1(J) = \sum_{j=0}^{J-1} \mu_j \quad (13)$$

$$\Sigma_2(J) = \sum_{j=0}^{J-1} \mu_j^2 \quad (14)$$

to obtain

$$\phi_j = \theta_j \quad (15)$$

$$\lambda_j = \lambda_j(J) = \frac{1}{J}\left[1 + \frac{J\mu_j - \Sigma_1(J)}{\sqrt{\frac{\Sigma_2(J) - \frac{1}{J}\Sigma_1^2(J)}{r - \frac{1}{J}}}}\right] \quad (16)$$

which hold for $j = 0, 1, \ldots, J-1$ where $J$ is defined by

$$J = \max\left\{j \,\middle|\, \lambda_{j-1}(j) > 0\right\} \quad (17)$$

By definition, all the eigenvalues $\lambda_j$ for $j \geq J$ are null.

*A. Finite dimensional signals*

The special case in which the signal to be acquired can be written as a linear combination of known waveforms by means of random coefficients is, for us, extremely interesting and deserves some further discussion.

Let us assume that (1) holds for orthonormal $u_j$ ($j = 0, \ldots, N-1$) and let us compute

$$C_a(t,s) = \sum_{j=0}^{N-1}\sum_{k=0}^{N-1} \mathbf{E}[a_j^* a_k] u_j^*(t) u_k(s) \quad (18)$$

The correlation matrix $\underline{A} = [\underline{A}_{j,k}] = [\mathbf{E}[a_j^* a_k]]$ is Hermitian and positive semidefinite, hence it can be written as $\underline{A} = \underline{Q}\underline{M}\underline{Q}^\dagger$ where $\cdot^\dagger$ stands for transposition and conjugation, $\underline{M}$ is a diagonal matrix with real non-negative diagonal entries, and $\underline{Q}$ is an orthonormal matrix whose columns are the eigenvectors of $\underline{A}$.

With this, we may rewrite (18) as

$$\begin{aligned} C_a(t,s) &= \sum_{j=0}^{N-1}\sum_{k=0}^{N-1}\sum_{l=0}^{N-1} \underline{Q}_{j,l}\underline{M}_{l,l}\underline{Q}_{k,l}^* u_j^*(s) u_k(t) \\ &= \sum_{l=0}^{N-1} \underline{M}_{l,l} \sum_{j=0}^{N-1} \underline{Q}_{j,l} u_j^*(t) \sum_{k=0}^{N-1} \underline{Q}_{k,l}^* u_k(s) \end{aligned}$$

Hence, we may express $C_a(t,s)$ in the form needed for writing (10) and thus the solution of (12) by simply setting $\theta_j = \sum_{k=0}^{N-1} \underline{Q}_{k,j}^* u_k$ and $\mu_j = \underline{M}_{j,j}$ for $j = 0, \ldots, N-1$.

This straightforward derivation clarifies that, when we have identified sparseness along a certain signal basis, the statistic of the coefficients gives us hints on the basis that may be used to highlight localization. Along this other basis, localization itself is nothing but the difference between the lower-index, largest eigenvalues $\mu_0, \mu_1, \ldots$ and the others.

A bridge is also built between the general treatment of rakeness in this Section and the frequency-domain analysis of the Section V. In fact, if $a$ is substantially bandlimited in the frequency interval $[-B, B]$ and is considered in the time interval $[-T/2, T/2]$ its realizations may be well expressed as a linear combination of waveforms that are the truncated version of prolate spheroidal wave functions [17][18]. It is known that, if $c = BT$ then $N = 2c$ functions are enough to achieve an approximation quality that dramatically increases as $c \to \infty$. Hence, the solution of (12) will feature $J = N = 2c$ for values of $r \in [1/J, 1] = [\frac{1}{2c}, 1]$.

From an operative point of view, whatever analysis allows us to obtain the generalized spectral expansion of $C_a$ as in (10), we may use (15), (16), (17) and (11) to compute the correlation function $C_b$ of the process generating the projection waveforms.

To fit this $C_b$ into an actual RMPI architecture, we must generate a binary or antipodal PAM signal with such a non-stationary correlation. The details of the mechanism allowing this are far beyond the scope of this paper and will be the topic of a future communication.

It is here enough to say that, if the number of symbols $S$ in each waveform is limited to few tens (say $S < 100$), we are able, depending on $C_b$, to automatically determine two sets of cardinality $s = S(S+1)/2$: the first set $\{z_0, z_1, \ldots, z_{s-1}\}$ contains sequences of modulating symbols, while the second set $\{\zeta_0, \zeta_1, \ldots, \zeta_{s-1}\}$ contains probabilities, so that $\sum_{j=0}^{s-1} \zeta_j = 1$.

These two sets are such that, if each time a projection waveform is needed, the modulating symbols in $z_j$ are used with probability $\zeta_j$, then the resulting process has the desired correlation.

In any case, let us stress that, as noted before for frequency-localized signals, the core of rakeness-based design concerns the solution of (5), which is here described for generically localized signals. Once that the correlation of the best projection waveforms is determined, their actual realization depends on implementation assumption that may vary from application to application.

## VII. SAMPLE APPLICATIONS

In this Section we introduce rakeness as a design criterion to optimize the performance of two acquisition systems, one that deals with Electro Cardio Graphic (ECG) signals, which can be easily modeled in the frequency domain as we did in Section V, and the other that must be described relying on the generalized spectral expansions in Section VI since its target signals are images.

Despite the fact that the two scenarios are different, the path we follow in designing an acquisition system based on CS system is the same and can be summarized in few steps:

i) identify the basis with respect to which the signal to acquire is sparse;
ii) identify the basis with respect to which the signal is localized;
iii) solve (5) for a number of possible values $r$ in its range;
iv) for each value of $r$, implement an RMPI architecture exploiting the sparsity revealed in i) and in which the projection waveforms are as close as possible to the optimal ones;
v) perform Monte-Carlo simulations to evaluate the resulting systems and select the best performing one.

Note that, in the classical design flow of a CS system, i) is a prerequisite while iv) is tackled once assuming that the projection waveform are random PAM signals with independently and identically distributed ("i.i.d." from now on) symbols. This is what will be taken as the reference case to quantitatively assess the improvements due to rakeness-based design.

In all cases, the performance index is the average reconstruction SNR (ARSNR), i.e. the average ratio between the energy of the original signal over the energy of the difference between the original signal and the reconstructed one. ARSNR values are always plotted at the center of an interval accounting for the variances of the corresponding reconstruction SNRs.

Note also that the implementation constraints (e.g., the restriction of projection waveform to PAM profiles with antipodal symbols) come into play only in iv).

Finally, one may observe that steps iii-v are nothing but an elementary line-search for the best possible value of $r$. As a matter of fact, the values of $r$ for which a definite improvement can be obtained are easily identified by means of a very small numbers of trials.

### A. Acquisition of ECGs

The ECG time shape represents the voltage between two different electrodes placed on the body at two specific positions. It records the electrical field produced by the myocardium, i.e., for each heart beat it cyclically reports the successive atrial depolarization/repolarization and ventricular depolarization/repolarization.

The application of CS techniques to ECG acquisition has been the topic of recent contributions [19][20][21][22] aimed to either reducing the amount of data needed to represent the signal in mobile applications or to achieve an high compression ratio in data storage systems.

In order to demonstrate the effectiveness of rakeness-based design for CS of ECGs, we need a broad collection of realistic realizations. To achieve this goal, we used a synthetic generator of ECGs, thoroughly discussed in [23] that provides signals not corrupted by noise to which we add white Gaussian noise with suitable power. The amount of noise is chosen so that the considered environment is realistic, but it can be arbitrary from the point of view of rakeness-based design that is independent of the noise level.

The generator core is expressed by the following set of three coupled ordinary differentially equations [23]

$$\begin{cases} \dot{x}_1 = \omega_1 x_1 - \omega_2 x_2 \\ \dot{x}_2 = \omega_1 x_1 + \omega_2 x_2 \\ \dot{x}_3 = -\sum_{i \in \{P,Q,R,S,T\}} \gamma_i \Theta_i \exp\left(-\frac{\Theta_i^2}{2v_i^2}\right) - (x_3 - \bar{x}_3) \end{cases} \quad (19)$$

Each heart beat is represented by a complete revolution on an attracting limit cycle in the $(x_1, x_2)$ plane. The shape of ECG signal is obtained introducing five attractors/repellors points in the $x_3$ direction in correspondence to the peaks and valleys that characterize the time shape of the signal and which are conventionally labeled by P, Q, R, S and T; furthermore $\bar{x}_3$ in (19) represents the mean value of the generated ECG.

TABLE I
PARAMETERS BOUNDS USED IN THE ECG GENERATOR.

| Index | P | Q | R | S | T |
|---|---|---|---|---|---|
| $\Theta_i$ | -75;-65 | -20;-5 | -5;5 | 10;20 | 95;105 |
| $\gamma_i$ | 1;1.4 | -5.2;-4.8 | 27;33 | -7.7;-7.3 | 0.5;1 |
| $v_i$ | 0.05;0.45 | -0.1;0.3 | -0.1;0.3 | -0.1;0.3 | 0.2;0.6 |

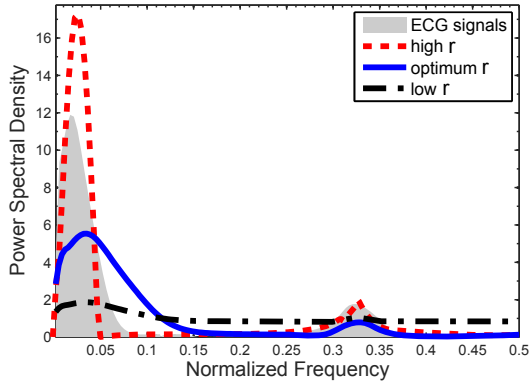

Fig. 4. Average spectra of real ECG signals (gray area) and of the sampling PAM sequences corresponding to the optimum (solid line) as well as an high (dashed line) and a low value (dash-dotted line) of $r$.

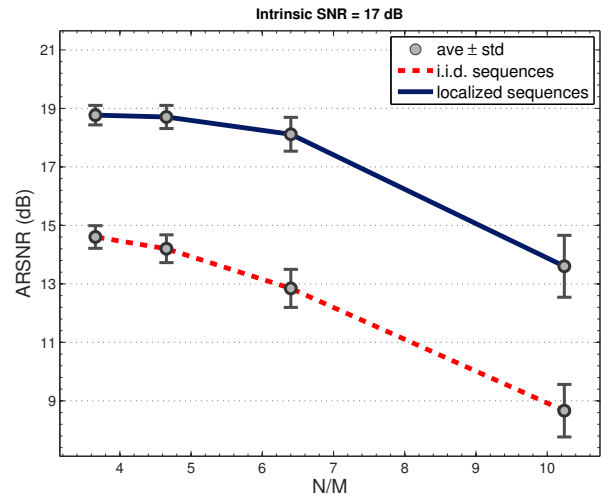

Fig. 5. Average value of the reconstructed SNR (ARSNR) as a function of the signal compression ratio N/M between the number of Nyquist samples and of CS measures. The dashed line refers to i.i.d. sampling waveforms and the solid line to rakeness-optimized ones.

In order to mimic the behavior of ECGs in patients affected by the most studied cardiac illness, the parameters $\gamma_i$, $v_i$ and $\Theta_i$, $i \in \{P,Q,R,S,T\}$, characterizing each considered signal are taken from a set of random variables uniformly distributed within the bounds reported in Table I. In addition, we randomly set the heart rate between 50Hz and 100Hz by property adjusting $\omega_1$ and $\omega_2$.

Though CS methods are classically developed for sparse representation with respect to signal bases, they have a straightforward generalizzation to sparse representation with respect to dictionaries, i.e., redundant collections of non-indipendent waveforms [24][25].

This is, in fact, the case of ECGs, for which a dictionary made of Gabor atoms

$$g_{s,u,v,w}(t) = \frac{1}{\sqrt{s}} e^{-\pi\left(\frac{t-u}{s}\right)^2} \cos(vt + w)$$

can be used [26][27].

In our experiment, a total of 507 atoms are used corresponding to different quadruple of parameters $(s, u, v, w)$. This collection of Gabor atoms is obtained using a greedy algorithm able to extract a limited number of functions from a broader set [27]. With respect to this dictionary the sparse representation of a typical ECG heartbeat waveform requires about 14 non-zero coefficients.

Furthermore, in our simulations, $T = 1$s within which the signal is sampled $N = 256$ times (a common choice for ECG equipments).

To apply the results in Section V, we first compute the average of the power spectral densities of 1000 ECGs generated as presented above to obtain $\hat{a}(f)$, the input of the optimization problem (8). The shape of $\hat{a}(f)$ is the gray profile shown in Figure 4. Next, we find the optimum $r$ as described at the beginning of Section VII. Figure 4 shows the optimum profile for the best value $r = 0.038$ (solid line) as well as for a smaller value (dash-dotted line) and for a larger value (dashed line).

Finally the LPF generator mentioned in Section V is used to produce the antipodal sequences with the optimized spectral profiles. These sequences are used to take $M$ measurements in a time window of length T, to which we add white Gaussian noise to construct the measurement vector $\underline{m}$ according to (2).

To determine the performance of rakeness based design, we consider a test set of 2000 ECG signals different from those employed for determining the average spectrum of Figure 4. These signals are acquired by projecting them both on localized antipodal sequences and on i.i.d. antipodal sequences (classically employed in CS-based methods and our reference case). The resulting ARSNRs are shown in Figure 5 as a function of the ratio between the intrinsic dimension of the signal and the number of CS measures. In both cases the intrinsic SNR is equal to 17dB.

As it can be noticed, rakeness-based design allows to achieve an improvement of at least 3.5dB in ARSNR with respect to the i.i.d. case, and even yields denoising (i.e. and ARSNR larger than the intrinsic SNR) for small compression ratio values. To give a visual representation of the improvement, Figure 6 reports, for $M = 32$, a comparison between an ECG signal and the reconstructed one for the i.i.d. (a) and rakeness-based (b) case. Direct visual inspection is enough to confirm the superiority of our approach.

### B. Acquisition of small images

The signal to acquire is a $24 \times 24$-pixel image, each pixel value ranging from 0 (black) to 1 (white), which represents a small white printed number or letter on a black background with a gray-level dithering to make the curves smoother to the human eye. Number and letters are randomly rotated and offset from the center of the image but never clipped.

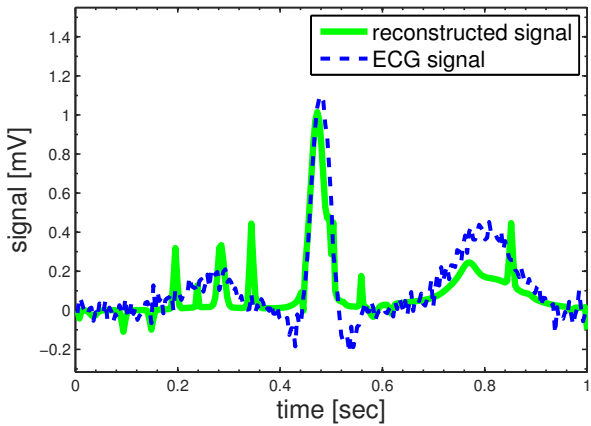

(a)

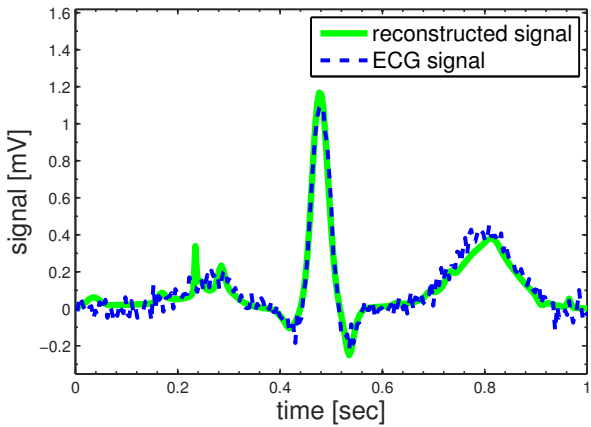

(b)

Fig. 6. Original (solid line) and reconstructed (dashed line) ECG when i.i.d sampling waveforms are used (plot (a)) and when rakeness-optimized sequence are exploited (plot (b)). In both cases N=256 and M=32 and the intrinsic SNR=17dB.

Although due to random rotations and offsets almost all pixels have a non-vanishing probability of being non-zero, a typical image contains only about 85 bright pixels, so that can be considered sparse in the base of 2-dim discrete delta functions that evaluate to 1 at a single pixel position and zero elsewhere. We may thus think of acquiring them using a RMPI architecture that projects along $24 \times 24$ antipodal random grids to obtain measurements that are enough to reconstruct the image but whose number $M$ is much less than the number $N = 24 \times 24 = 576$ of the original pixels.

To simplify the design phase, the generation of the random grids is done by adjoining $4 \times 4$ subgrids each with $6 \times 6$ antipodal values whose statistic is optimized by solving (12).

To allow calculations, the values in each subgrid are rearranged into a 36-dimensional vector as schematically reported in Figure 7, that also highlights the subgrid on which we will focus in the following.

In that region, and due to the vector rearrangement, we may list the modulating symbols of the projection waveform $b$ with $b_j$ for $j = 0, \ldots, 35$. The same can be done for the incoming signal $a$ when it is expressed along the basis of 2-dim discrete

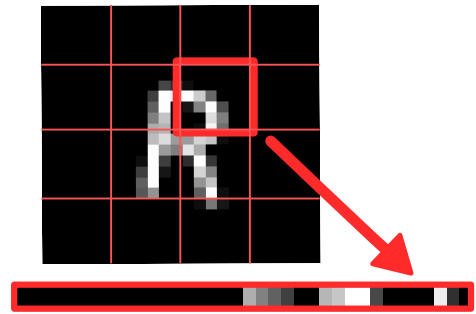

Fig. 7. A sample image, its partition and rearrangement into a vector containing the value of each pixel.

delta's with coefficients that we may indicate with $a_j$ for $j = 0, \ldots, 35$.

If $\underline{a} = (a_0, \ldots, a_{35})^\top$, we may follow the development of subsection VI-A to estimate the $36 \times 36$ matrix $\underline{A} = \mathbf{E}[\underline{a}\underline{a}^\top]$ by empirical averaging over a training set of 2080 randomly generated images.

The resulting matrix is reported in graphic form in Figure 8-(a) where, for each pair of indexes $j, k = 0, \ldots, 35$, a point is laid down whose brightness is proportional to the values of $\underline{A}_{j,k} = \mathbf{E}[a_j a_k]$.

The eigenvalues $\mu_0, \ldots, \mu_{35}$ of that $\underline{A}$ are reported as the light bars in Figure 8-(c). By exploiting (16) and (17) for $r = 0.047$ we get $J = 36$ and the eigenvalues $\lambda_0, \ldots, \lambda_{35}$ reported as dark bars in figure 8-(c).

From the eigenvectors of $\underline{A}$ and these new eigenvalues, we may construct the correlation matrix of the values in this projection subgrid. Since the subgrid contains $6 \times 6 = 36$ values, its correlation matrix $\underline{B}$ has dimensions $36 \times 36$ and is reported in Figure 8-(b) in a graphical form adopting the same convention used to represent the values of $\underline{A}$.

Once that $\underline{B}$ is known, we are able to select a collection of $36 \times 35/2 = 630$ antipodal subgrids $z_0, \ldots, z_{629}$ with attached probabilities $\zeta_0, \ldots, \zeta_{629}$ so that, if each time a projection is needed we use the subgrid $z_j$ with probability $\zeta_j$, the overall process features exactly that correlation matrix.

The same design process is repeated for each of the 4 central $6 \times 6$ regions in the image while the 12 outer subgrids are built from independent and uniformly distributed antipodal symbols. All the subgrids are finally compounded in a complete $24 \times 24$ projection grid.

As a comparison case, projections are also taken by using i.i.d. symbols for all the elements of the projection grid.

In both cases, noise is added to the projections before they take their place in the vector $\underline{m}$ of $M$ measurements according to (2), and reconstruction is performed using the algorithm reported in [9].

Figure 9 reports the ARSNR (over 3000 trials) of the reconstructed images and compares the performance of an RMPI based on rakeness-optimized projections and on i.i.d. projections for different values of the compression ratio $N/M$. In both cases the intrinsic SNR is 17dB.

It is evident from Figure 9 that, even if it is exploited only in the central portion of the images, rakeness-based design

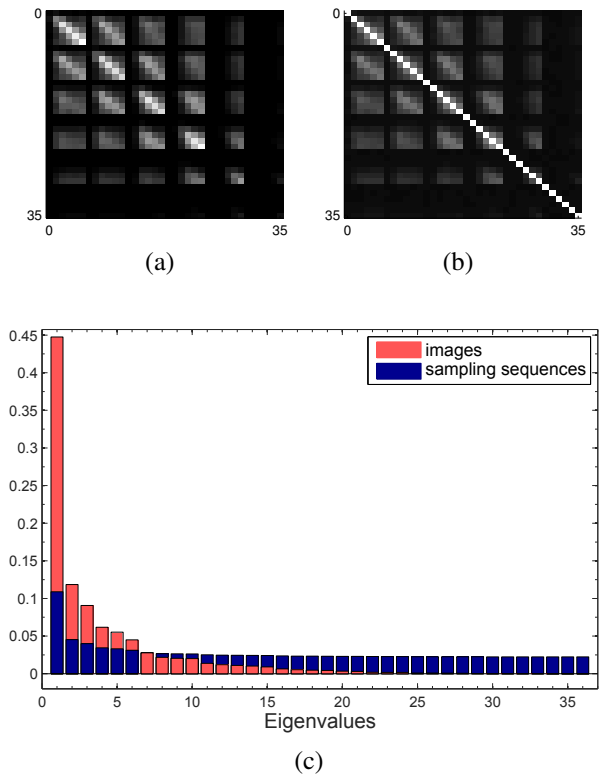

Fig. 8. Correlation matrix of the pixels in one of the four central regions (a) Correlation matrix of the optimal projection processs (b) Eigenvalues of the above correlation matrices (c).

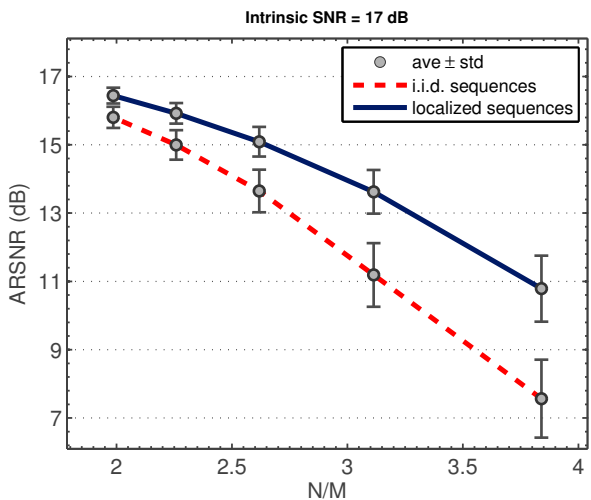

Fig. 9. Quality of the reconstructed images when rakeness-optimized or i.i.d. projection grids are used in an RMPI architecture for different compression ratios.

leads to non-negligible improvement of at least 1dB.

A qualitative appreciation of such an improvement can be obtained from Figure 10 in which 5 images (a) are acquired and reconstructed by means of $M = 115$ rakeness-optimized projections (b) or by the same number of i.i.d. projections (c). Reconstruction artifacts are visibly reduced by adopting rakeness-based design.

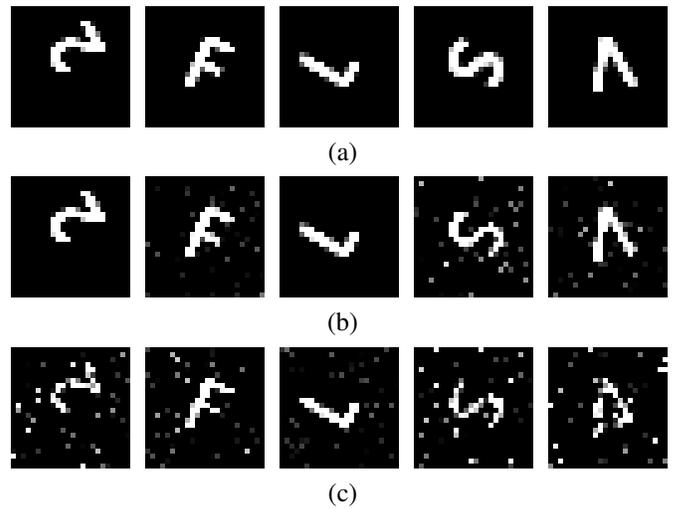

Fig. 10. Sample images (a) and their reconstruction based on rakeness-optimized projection grids (b) or on i.i.d. projection grids (c).

## VIII. CONCLUSION

Compressive sensing exploits the fact that, when looked at in the right domain, the information content of a signal can be much less than what appears when we look at it in time or frequency (i.e., the signal is sparse).

Acquisition schemes that exploit sparsity may lead to considerable advantages in terms of sensing system design since, for example, if the information content is much less than the signal bandwidth, sub-Nyquist sampling can be employed.

To all this we add the consideration that, in a possibly different domain, the energy of the signals may be not uniformly distributed (i.e., the signal is localized) and when noise is present, it is convenient to adapt the system to "rake" as much signal energy as possible.

By itself, this is not a novel concept since it appears, for example, in matched filters and rake receivers used in telecommunications. Yet, in our context, the efforts to collect the energy of the signal must be balanced with the guarantee that all details of its underlying structure can be captured when immersed in noise. This brings us to a trade-off that we propose to address in statistical terms by means of an optimization problem: maximize the "rakeness" while obeying to a constraint ensuring that the measurements are random enough to capture all signal details.

The paper develops the formal definition of such problem as well as its solution for stationary signals whose localization can be highlighted in the frequency domain, and for more generic non-stationary signals whose localization is more evident in suitably defined domains.

The applicability of both techniques is demonstrated by sample applications to the acquisition of ECG tracks and small letter images.

## IX. APPENDIX

*A. Solution of* (12)

The first subproblem $\max_\Xi$ can be solved leveraging on the fact that it is a linear problems with linear constraints. Since,

in principle, it may involve an infinite number of variables we should proceed by steps.

Let $P_n$ be the optimization problem

$$\max_\Xi \sum_{j=0}^{n-1} \sum_{k=0}^{n-1} \lambda_j \mu_k \Xi_{j,k}$$
$$\text{s.t.} \quad \begin{array}{rcll} \Xi_{j,k} & \geq & 0 & \forall j,k \\ \sum_{j=0}^{\infty} \Xi_{j,k} & = & 1 & \forall k \\ \sum_{k=0}^{\infty} \Xi_{j,k} & = & 1 & \forall j \end{array}$$

so that $P_\infty$ is the basis finding subproblem in (12).

Since all the series involved in the definition of $P_\infty$ are convergent, we have that, independently of $\Xi_{j,k}$,

$$\lim_{n \to \infty} \sum_{j=0}^{n-1} \sum_{k=0}^{n-1} \lambda_j \mu_k \Xi_{j,k} = \sum_{j=0}^{\infty} \sum_{k=0}^{\infty} \lambda_j \mu_k \Xi_{j,k}$$

Moreover, since all the summands are positive, the limit is from below.

Let us now assume to have solved $P_\infty$ yielding a value $\sigma(P_\infty)$ corresponding to a certain optimal choice $\hat{\Xi}_{j,k}^\infty$.

Given any $\epsilon > 0$ there is a $\bar{n}$ such that for any $n \geq \bar{n}$

$$0 \leq \sigma(P_\infty) - \sum_{j=0}^{n-1} \sum_{k=0}^{n-1} \lambda_j \mu_k \hat{\Xi}_{j,k}^\infty \leq \epsilon$$

Yet, by solving $P_n$ we get a solution $\sigma(P_n)$ such that

$$\sum_{j=0}^{n-1} \sum_{k=0}^{n-1} \lambda_j \mu_k \hat{\Xi}_{j,k}^\infty \leq \sigma(P_n) \leq \sigma(P_\infty)$$

where the last inequality holds since every feasible configuration for $P_n$ is also a feasible configuration for $P_\infty$.

Altogether we get that for any $n \geq \bar{n}$

$$0 \leq \sigma(P_\infty) - \sigma(P_n) \leq \epsilon$$

that is

$$\lim_{n \to \infty} \sigma(P_n) = \sigma(P_\infty)$$

from below.

From this we know that, if the solutions $\hat{\Xi}_{j,k}^n$ of $P_n$ have a limit, such a limit yields $\sigma(P_\infty)$.

To study the solutions of $P_n$ we may first recall that the polytope

$$\begin{array}{rcll} \Xi_{j,k} & \geq & 0 & j,k=0,\ldots,n-1 \\ \sum_{j=0}^{n-1} \Xi_{j,k} & = & 1 & k=0,\ldots,n-1 \\ \sum_{k=0}^{n-1} \Xi_{j,k} & = & 1 & j=0,\ldots,n-1 \end{array}$$

is the one characterizing the so-called "assignment" problems [28] and is well known [29] to have vertices for $\Xi_{j,k}$ for $j,k = 0,\ldots,n-1$ equal to a permutation matrix. Hence, let $\xi : \{0,1,\ldots,n-1\} \mapsto \{0,1,\ldots,n-1\}$ be the bijection such that

$$\Xi_{j,k} = \begin{cases} 1 & \text{if } k = \xi(j) \\ 0 & \text{otherwise} \end{cases}$$

we have

$$\sigma(P_n) = \sum_{j=0}^{n-1} \lambda_j \mu_{\xi(j)}$$

for some optimally chosen $\xi$.

Actually, we may prove that such an optimal $\xi$ is the identity. We do it by induction.

For $n = 2$ there are only two permutations corresponding to the two candidate solutions $\sigma' = \lambda_0 \mu_0 + \lambda_1 \mu_1$ and $\sigma'' = \lambda_0 \mu_1 + \lambda_1 \mu_0$. Yet, from the sorting of the $\lambda_j$ and of the $\mu_j$ we have $\sigma' - \sigma'' = (\lambda_0 - \lambda_1)(\mu_0 - \mu_1) \geq 0$.

This confirms that the optimum solution is the one corresponding to $\xi(j) = j$ for $j = 0, 1$.

Assume now that this is true for $n$ up to a certain $\bar{n}$ and that we have solved $P_{\bar{n}+1}$ by means of a permutation $\xi$.

If $\xi(0) = \bar{j} > 0$ then $\sigma(P_{\bar{n}+1}) = \lambda_0 \mu_{\bar{j}} + \sigma'$. Yet, $\sigma'$ must be the value of the solution of a problem with $\bar{n}$ terms $\lambda_1, \ldots, \lambda_{\bar{n}}$ and $\mu_1, \ldots, \mu_{\bar{j}-1}, \mu_{\bar{j}+1}, \ldots, \mu_{\bar{n}}$. Since we assumed to know how problems with $\bar{n}$ terms are solved we know that

$$\sigma' = \sum_{j=1}^{\bar{j}} \lambda_j \mu_{j-1} + \sum_{j=\bar{j}+1}^{\bar{n}} \lambda_j \mu_j$$

It is now easy to see that the value $\lambda_0 \mu_{\bar{j}} + \sigma'$ of the alleged solution is actually smaller than $\sum_{j=0}^{\bar{n}} \lambda_j \mu_j$.

In fact

$$\begin{aligned}
\sum_{j=0}^{\bar{n}} \lambda_j \mu_j - \lambda_0 \mu_{\bar{j}} - \sum_{j=1}^{\bar{j}} \lambda_j \mu_{j-1} - \sum_{j=\bar{j}+1}^{\bar{n}} \lambda_j \mu_j &= \\
= \lambda_0(\mu_0 - \mu_{\bar{j}}) - \sum_{j=1}^{\bar{j}} \lambda_j(\mu_{j-1} - \mu_j) & \\
= \lambda_0 \sum_{j=1}^{\bar{j}}(\mu_{j-1} - \mu_j) - \sum_{j=1}^{\bar{j}} \lambda_j(\mu_j - \mu_{j-1}) & \\
= \sum_{j=1}^{\bar{j}}(\lambda_0 - \lambda_j)(\mu_{j-1} - \mu_j) \geq 0 &
\end{aligned}$$

Hence, the optimal permutation must feature $\xi(0) = 0$. This reduces the solution of $P_{\bar{n}+1}$ to the solution of $P_{\bar{n}}$ that we already know to be $\xi(j) = j$ for $j = 1, \ldots, n-1$.

In the light of this, every $P_n$ has a solution corresponding to $\xi(j) = j$ for $j = 0, \ldots, n-1$ and the solution of $P_\infty$ is

$$\Xi_{j,k} = \delta_{j,k}$$
$$\sigma(P_\infty) = \sum_{j=0}^{\infty} \lambda_j \mu_j$$

This solves the basis-selection problem and yields (15). The original (12) now becomes

$$\max_\lambda \sum_{j=0}^{\infty} \lambda_j \mu_j$$
$$\text{s.t.} \quad \begin{array}{rcl} \lambda_j & \geq & 0 \quad \forall j \\ \sum_{j=0}^{\infty} \lambda_j & = & 1 \\ \sum_{j=0}^{\infty} \lambda_j^2 & \leq & r \end{array} \quad (20)$$

Since the $\lambda_j$ are non-negative and sorted in non-increasing order we have that the set of indexes such that $\lambda_j > 0$ must be of the kind $\{0, 1, \ldots, J-1\}$ for some integer $J \geq 0$. We also know that, to allow $\sum_{j=0}^{J-1} \lambda_j = 1$ and $\sum_{j=0}^{J-1} \lambda_j^2 = r$ to hold simultaneously we must have $r \geq 1/J$ and thus $J \geq 1/r$.

Hence, for a given $J \geq 1/r$ our problem can be recast into

$$\max_\lambda \quad \sum_{j=0}^{J-1} \lambda_j \mu_j$$
$$\text{s.t.} \quad \begin{array}{rcl} \lambda_j & > & 0 \quad j = 0, \ldots, J-1 \\ \sum_{j=0}^{J-1} \lambda_j & = & 1 \\ \sum_{j=0}^{J-1} \lambda_j^2 & \leq & r \end{array} \quad (21)$$

Note that the feasibility set of (21) for a certain $J = \bar{J}$ contains points that are arbitrarily close to those of the feasibility set of (21) for any $J < \bar{J}$. Hence, to maximize the rakeness we should try to have $J$ as large as possible.

To determine the $J$ leading to maximum rakeness note first that, if we drop the randomness constraint $\sum_{j=0}^{J-1} \lambda_j^2 \leq r$, the relaxed problem has the trivial solution $\lambda_0 = 1$ and $\lambda_j = 0$ for $j > 0$. Such a solution is not feasible for the original problem since $\sum_{j=0}^{J-1} \lambda_j^2 = 1 \geq r$, hence the corresponding optimum must be attained when the randomness constraint is active, i.e. for $\sum_{j=0}^{J-1} \lambda_j^2 = r$.

The Karush-Kuhn-Tucker conditions for (20) with the inequality constraint substituted by the equality constraint are

$$\begin{array}{rcl} \mu_j + \ell' + \ell''\lambda_j + \ell'''_j & = & 0 \quad \forall j \\ \lambda_j & \geq & 0 \quad \forall j \\ \sum_{j=0}^{\infty} \lambda_j & = & 1 \\ \sum_{j=0}^{\infty} \lambda_j^2 & = & r \\ \ell'''_j & \geq & 0 \quad \forall j \\ \ell'''_j \lambda_j & = & 0 \quad \forall j \end{array}$$

where $\ell'$ is the Lagrange multiplier corresponding to $\sum_{j=0}^{\infty} \lambda_j = 1$, $\ell''$ is the Lagrange multiplier corresponding to $\sum_{j=0}^{\infty} \lambda_j^2 = r$, and $\ell'''_j$ are the Lagrange multipliers corresponding to $\lambda_j \geq 0$ which must hold $\forall j$.

Since for $\lambda_j > 0$ the constraint $\ell'''_j \lambda_j = 0$ sets $\ell'''_j = 0$ we know that

$$\lambda_j = -\frac{\mu_j + \ell'}{\ell''} \quad (22)$$

for $j = 0, \ldots, J-1$.

Since the sequences $\lambda_j$ and $\mu_j$ are both decreasing, we must have $\ell'' < 0$ and thus $\ell' > -\mu_j$ for $j = 0, 1, \ldots, J-1$, i.e., $\ell' > -\mu_{J-1}$. Hence,

$$J = \max\{j \mid \ell' > -\mu_{j-1}\}$$

To see how the two parameters $\ell'$ and $\ell''$ depend on $J$ note that they should satisfy the simultaneous equations

$$\begin{array}{rcl} \sum_{j=0}^{J-1} \frac{\mu_j + \ell'}{-\ell''} & = & 1 \\ \sum_{j=0}^{J-1} \left(\frac{\mu_j + \ell'}{\ell''}\right)^2 & = & r \end{array}$$

i.e., exploiting (13) and (14),

$$\begin{array}{rcl} \Sigma_1(J) + J\ell' & = & -\ell'' \\ \Sigma_2(J) + 2\Sigma_1(J)\ell' + J\ell'^2 & = & \ell''^2 r \end{array}$$

Such equations can be solved for $\ell'$ and $\ell''$ and the resulting values substituted in (22) to yield (16).

### B. Real and nonnegative values in (16)

The denominator within the square root is positive whenever $r > 1/J$.

To show that the corresponding numerator is also non-negative write

$$J\Sigma_2(J) - \Sigma_1^2(J) =$$
$$= J\sum_{j=0}^{J-1} \mu_j^2 - \sum_{j=0}^{J-1} \sum_{k=0}^{J-1} \mu_j \mu_k$$
$$= J\sum_{j=0}^{J-1} \mu_j \left[\mu_j - \frac{1}{J}\sum_{k=0}^{J-1} \mu_k\right]$$

Let now $\zeta_j = \mu_j - \frac{1}{J}\sum_{k=0}^{J-1} \mu_k$. We have that the $\zeta_j$ are decreasing and such that $\sum_{j=0}^{J-1} \zeta_j = 0$. Hence, there is a $j'$ such that $\sum_{j=0}^{j'-1} \zeta_j = \sum_{j=j'}^{J-1}(-\zeta_j) \geq 0$. Hence,

$$J\Sigma_2(J) - \Sigma_1^2(J) =$$
$$= J\left[\sum_{j=0}^{j'-1} \mu_j \zeta_j - \sum_{j=j'}^{J-1} \mu_j(-\zeta_j)\right]$$
$$\geq J\left[\mu_{j'-1}\sum_{j=0}^{j'-1} \zeta_j - \mu_{j'}\sum_{j=j'}^{J-1}(-\zeta_j)\right]$$
$$= J(\mu_{j'-1} - \mu_{j'})\sum_{j=0}^{j'-1} \zeta_j \geq 0$$

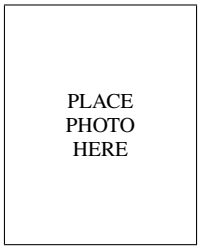

**Mauro Mangia** (S'10) He was born in Lecce, Italy. He received the B.S. and M.S. degree in electronic engineering from the University of Bologna, Italy, in 2004 and 2009 respectively. He is currently a PhD Student in the Information Techology under the European Doctorate Project (EDITH) from University of Bologna, Italy. In 2009 he was a Visiting Scholar at Non-Linear System Laboratory of the École Polytechnique Fédérale de Lausanne (EPFL). His research interests are in non linear systems, compressed sensing, ultra wide band systems and system biology. He was recipient of the best student paper award at ISCAS2011.

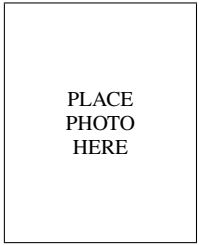

**Riccardo Rovatti** (M'99, SM'02, F'12) received the M.S. degree (summa cum laude) in electronic engineering and the Ph.D. degree in electronics, computer science, and telecommunications from the University of Bologna, Italy, in 1992 and 1996, respectively.

Since 2001 he is Associate Professor of Electronics with the University of Bologna. He is the author of more than 230 technical contributions to international conferences and journals and of two volumes. He is co-editor of the book Chaotic Electronics in Telecommunications (CRC, Boca Raton) as well as one of the guest editors of the May 2002 special issue of the PROCEEDINGS OF THE IEEE on "Applications of Non-linear Dynamics to Electronic and Information Engineering." His research focuses on mathematical and applicative aspects of statistical signal processing especially those concerned with nonlinear dynamical systems. Prof. Rovatti was an Associated Editor of the IEEE TRANSACTIONS ON CIRCUITS AND SYSTEMS–PART I. In 2004 he received the Darlington Award of the Circuits and Systems Society. He was the Technical Program Cochair of NDES 2000 (Catania) and the Special Sessions Cochair of NOLTA 2006 (Bologna).

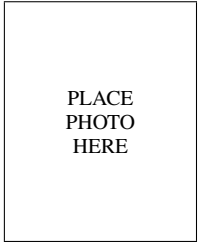

**Gianluca Setti** (S'89, M'91, SM'02, F'06) received a Dr. Eng. degree (with honors) in Electronic Engineering and a Ph.D. degree in Electronic Engineering and Computer Science from the University of Bologna, Bologna in 1992 and in 1997, respectively, for his contribution to the study of neural networks and chaotic systems. From May 1994 to July 1995 he was with the Laboratory of Nonlinear Systems (LANOS) of the Swiss Federal Institute of Technology in Lausanne (EPFL) as visiting researcher. Since 1997 he has been with the School of Engineering at the University of Ferrara, Italy, where he is currently a Professor of Circuit Theory and Analog Electronics. He held several visiting position at Visiting Professor/Scientist at EPFL (2002, 2005), UCSD (2004), IBM T. J. Watson Laboratories (2004, 2007) and at the University of Washington in Seattle (2008, 2010) and is also a permanent faculty member of ARCES, University of Bologna. His research interests include nonlinear circuits, recurrent neural networks, implementation and application of chaotic circuits and systems, statistical signal processing, electromagnetic compatibility, wireless communications and sensor networks.

Dr. Setti received the 1998 Caianiello prize for the best Italian Ph.D. thesis on Neural Networks and he is co-recipient of the 2004 IEEE CAS Society Darlington Award, as well as of the best paper award at ECCTD2005 and the best student paper award at EMCZurich2005 and at ISCAS2010.

He served as an Associate Editor for the IEEE Transactions on Circuits and Systems - Part I (1999-2002 and 2002-2004) and for the IEEE Transactions on Circuits and Systems - Part II (2004-2007), the Deputy-Editor-in-Chief, for the IEEE Circuits and Systems Magazine (2004-2007) and as the Editor-in-Chief for the IEEE Transactions on Circuits and Systems - Part II (2006-2007) and of the IEEE Transactions on Circuits and Systems - Part I (2008-2009).

He was the 2004 Chair of the Technical Committee on Nonlinear Circuits and Systems of the of the IEEE CAS Society, a Distinguished Lecturer (2004-2005), a member of the Board of Governors (2005-2008), and he served as the 2010 President of the same society.

Dr. Setti was also the Technical Program Co-Chair of NDES2000 (Catania) the Track Chair for Nonlinear Circuits and Systems of ISCAS2004 (Vancouver), the Special Sessions Co-Chair of ISCAS2005 (Kobe) and ISCAS2006 (Kos), the Technical Program Co-Chair of ISCAS2007 (New Orleans) and ISCAS2008 (Seattle), as well as the General Co-Chair of NOLTA2006 (Bologna).

He is co-editor of the book *Chaotic Electronics in Telecommunications* (CRC Press, Boca Raton, 2000), *Circuits and Systems for Future Generation of Wireless Communications* (Springer, 2009) and *Design and Analysis of Biomolecular Circuits* (Springer, 2011), as well as one of the guest editors of the May 2002 special issue of the IEEE Proceedings on "Applications of Non-linear Dynamics to Electronic and Information Engineering".

He is a Fellow of the IEEE.